# On the generation of solar spicules and Alfvénic waves

J. Martínez-Sykora[1,2] *, B. De Pontieu[2,3], V. H. Hansteen[3,2], L. Rouppe van der Voort[3], M. Carlsson[3], T. M. D. Pereira[3]

[1]Bay Area Environmental Research Institute, Petaluma, CA 94952, USA.

[2]Lockheed Martin Solar and Astrophysics Laboratory (LMSAL), Palo Alto, CA 94304, USA.

[3]Institute of Theoretical Astrophysics, University of Oslo, P.O. Box 1029 Blindern, N-0315 Oslo, Norway

*E-mail: juanms@lmsal.com

In the lower solar atmosphere, the chromosphere is permeated by jets known as spicules, in which plasma is propelled at speeds of 50-150 km s$^{-1}$ into the corona. The origin of the spicules is poorly understood, although they are expected to play a role in heating the million-degree corona and are associated with Alfvén waves that help drive the solar wind. We compare magnetohydrodynamic simulations of spicules with observations from the Interface Region Imaging Spectrograph and the Swedish Solar Telescope. Spicules are shown to occur when magnetic tension is amplified and transported upwards through interactions between ions and neutrals or ambipolar diffusion. The tension is impulsively released to drive flows, heat plasma (through ambipolar diffusion) and generate Alfvén waves.

One Sentence Summary: Magnetic tension and ion-neutral interactions drive chromospheric jets linked to coronal heating and Alfvenic waves that help drive the solar wind.

Spicules are ubiquitous, highly dynamic jets of plasma that are observed at the solar limb*(1-3)*. Recent observations indicate that in most of these jets (type II spicules) initially cool (~10$^4$ K) plasma is accelerated into the corona at speeds of up to 150 km s$^{-1}$. A substantial fraction of the ejected plasma is heated to temperatures typical of the solar transition region (TR, >10$^4$ K *(3-5)*, this narrow region separates the cool chromosphere from the hot corona) before falling back to the surface after 5-10 minutes. Spicules may play a critical role in energizing the outer atmosphere and have been suggested as a source of hot plasma to the corona[*(6)*, but see *(7)*] potentially helping explain its puzzling temperatures of several million degrees. Spicules carry a large flux of Alfvénic waves*(8-10)* that may help drive the solar wind and/or heat the corona*(11)*. Despite major observational advances, the generation of type II spicules remains poorly understood: Although there are suggestions that magnetic reconnection plays a role*(6)*, there are no models at present *(12)* that can explain all of the observed properties of type II spicules including their ubiquity*(3-5)* and the strong magnetic waves they carry*(8,9,13)*.

We present 2.5-dimensional (2.5D) radiative magnetohydrodynamic (MHD) numerical simulations in which spicule-like features naturally and frequently occur. Our model captures the complex physical processes that play a role in spicule formation and evolution: (i) Plasma is not in local thermodynamic equilibrium, (ii) radiation is optically thick and undergoes scattering, (iii) gas is partially ionized, and (iv) thermal conduction is important in the upper chromosphere, TR and corona (Supplementary material). Although previous numerical experiments included many of these processes*(12)*, they did not produce any*(14)* or produced only a few*(15)* features that resembled type II spicules because they lacked ambipolar diffusion (Supplementary Text). We find that high spatial resolution (<40 km), large-scale magnetic field (e.g., magnetic loops ~50 Mm long), and ion-neutral interaction effects in the partially ionized chromosphere are critical for the ubiquitous formation of spicules. Our results show that spicule-like jets occur frequently in the vicinity of strong magnetic flux concentrations (e.g., the positions on the horizontal axis $x$~20, $x$~40 and $x$~70 Mm in Figure 1, movie S1) that are similar to so-called plage regions (large concentrations of magnetic flux with predominantly one polarity). We can identify at least two different drivers for type II spicules in the model. One (described below) is much more frequent than the other (described in the Supplementary material).

Several steps are required for the formation of spicule-like features. First, convective motions in the photosphere distort the magnetic field in the vicinity of strong vertical magnetic flux concentrations. This occurs through the interaction of these vertical concentrations with neighboring horizontal fields associated with weaker, granular-scale flux concentrations (left panels, Fig. 2). In an environment in which plasma β (the ratio of plasma to magnetic pressure) is high (i.e, gas pressure greater than magnetic pressure), this interaction can lead to the local build-up of strong magnetic tension. In our simulations such weak fields continuously appear as a result of processing and shredding of strong fields caused by convective motions. These conditions are likely to be found frequently on the Sun in the vicinity of regions that are dominated by mostly vertical magnetic flux concentrations of several kilogauss, such as magnetic network (in the quiescent Sun) or plage (in active regions). These are surrounded by weak, mostly horizontal flux that continuously emerges on granular scales of ~$10^3$ km*(16)*.

Next, this region of highly bent magnetic field and thus large magnetic tension must move into the upper chromosphere where the magnetic field dominates the plasma (β<1). However, weak horizontal fields are typically not buoyant enough to emerge into the atmosphere*(17)*. The process that allows this region of magnetic tension to move to greater heights is ambipolar diffusion, which arises from slippage between the ions and neutral particles in the partially ionized chromosphere. The ambipolar diffusion allows magnetic field to move through neutral particles, and the collisions between neutrals and ions can lead to magnetic energy dissipation. Ambipolar diffusion depends on the ion-neutral collision frequency and magnetic field strength, which is largest in the coldest parts of the chromosphere and affects the emergence of regions with high magnetic tension in two ways: (i) Sporadically, ambipolar diffusion becomes large just above the photosphere as a result of the strong expansion and adiabatic cooling in the wake of magneto-acoustic shocks that propagate upwards. When this happens close to a region of high magnetic tension, the ion-neutral interaction diffuses the mostly horizontal upper photospheric fields into the chromosphere, allowing the highly bent magnetic field to penetrate upwards into a low plasma β regime. (ii) The emergence into the chromosphere leads to expansion of the field, which in turns leads to cooling and a dramatic increase in ambipolar diffusion within the emerging region. As a result, any current perpendicular to the field lines (Fig. 2M-P, movie S2) is partially dissipated in the cold expanding pockets and partially advected to the sides of the cold regions, where the ambipolar diffusion is lower (Fig. 2I-L). This current leads to a further amplification of the magnetic tension and its concentration into a narrower layer at the boundaries of the emerging flux bubble.

In the final step, the magnetic tension is violently released in the upper chromosphere, which drives strong flows (~100 km s$^{-1}$, Fig. 2E-H). The magnetic field releases its tension by retracting and straightening, similar to the whiplash effect during magnetic reconnection*(18)*. The straightening field squeezes the plasma in a confined region, which produces, through pressure gradients, a strong acceleration of the chromospheric plasma to high speeds along the ambient magnetic field. The whiplash effect also generates Alfvén waves that rapidly propagate upward, as well as electrical currents.

Our simulation reproduces many observed properties of type II spicules: (Fig. 3, movie S3), including highly collimated and strong flows (~100 km s$^{-1}$) that reach heights up to 10 Mm*(4,19)* within a lifetime of 2-10 minutes*(5)*. The vertical motion of the TR above the spicule peaks at 30-35 km s$^{-1}$ which translates to an apparent motion along the spicule of ~40-45 km s$^{-1}$ (Figure 3F) when taking into account the inclination, in agreement with observed apparent motions along spicules at the solar limb*(4)*. Our model predicts spicules to occur in the vicinity of the network and plage, especially towards the periphery of strong flux regions where interactions with weaker flux can easily occur. Our simulation shows a much higher spicule occurrence rate than previous studies*(15)*, which produced only two examples. Although those two spicule-like features were also propelled by magnetic tension, the simulations lacked ambipolar diffusion and the underlying driver was actually injection of strong field at the lower boundary. These earlier 3D simulations did not reproduce many observed properties of spicules*(12)*. However, similar driving mechanism in the 3D simulations suggests that the 2.5D limitation of our simulations does not meaningfully affect results. This is also supported by the highly collimated nature of the flows and shocks in our simulations (discussed below and in the Supplementary Material).

Our model predicts heating from chromospheric (<10$^4$ K) to TR (>2x10$^4$ K) temperatures during the evolution of the spicules (Fig. 3, movie 3). Currents driven by the whiplash effect are partially dissipated (Fig. 2M-P) through ambipolar diffusion thereby heating the chromospheric spicular plasma to at least TR temperatures within 2 minutes during its expansion into the corona (Fig. 3S-X). This is compatible with observations that show a relatively short (<1.5 min), cool (visible in the Ca II lines) initial phase followed by a longer-lived phase (~5-10 min) at higher temperatures (visible in Mg II and Si IV lines)*(5)*. In the model, the remaining currents penetrate into the corona, are dissipated via Joule heating and the connected magnetic loops and spicule-associated plasma reach coronal temperatures of 2 MK (Figs. 2, 4, movie S2).

We generated synthetic observations of the spicules in our simulation for comparison with observations from the Swedish 1-m Solar Telescope (SST)*(20)* and NASA's Interface Region Imaging Spectrograph (IRIS)*(21)*. We compared the Ca II 8542 Å (middle chromosphere, ~8x$10^3$ K), Mg II h 2803 Å (upper chromosphere (~1-1.5x$10^4$ K), Si IV 1402 Å (~8x$10^4$ K) and Fe IX 171 Å (~1 MK) spectral lines (Fig. 4). The model reproduces the short lifetimes (<1 min) of high-velocity excursions in the blue wing of chromospheric lines such as Ca II 8542 Å and Mg II h 2803 Å, so-called rapid blueshifted events (RBEs) that are the disk counterparts of type II spicules*(22-23)*. The strong upward velocities at chromospheric temperatures (Fig. 3G-L, movie S3) last for only a few tens of seconds. The density along the spicule decreases with time (Fig. 3M-R, movie S3), which contributes to the short lifetime in chromospheric lines. The TR counterparts of the model spicules are visible as strong blueshifted excursions of the Si IV line, which agrees well with IRIS observations (Fig. 4C and 4G). The modeled spicule also presents a signal in the coronal Fe IX 171 Å line, similar to what has been observed before*(24)*.

Our simulation indicates that spicules may play a substantial role in energizing the outer solar atmosphere, by supplying plasma to the corona and by the generation of strong electrical currents and Alfvén waves. Assuming that the spicules have width of 300 km in all directions, a spicule supplies ~$10^{10}$ kg of hot plasma to the corona per event. The strong intermittent currents that are an integral driving component of spicules fill the spicule and propagate into the corona. These currents are dissipated by the ambipolar diffusion in chromospheric or spicular plasma which leads to substantial heating on the order of ~$10^{18}$ J per event (integrated over the spicule lifetime). Ambipolar diffusion is not effective at coronal temperatures, but these spicule-associated currents also have clear potential for heating of plasma in the coronal volume*(25-26)*. This is illustrated by the hot coronal loops that appear in association with the simulated spicules, which are filled and heated by current dissipation (in our simulation by artificial diffusion; Supplementary material), strong flows and shocks, evaporation, and thermal conduction.

Alfvénic motions have additional capacity for heating plasma to coronal temperatures*(11)*. The Alfvénic motions that are generated through the whiplash effect lead to transverse waves in our 2.5D simulations, and would lead to torsional and kink waves on the Sun. In our simulation the average amplitude for these waves is 20 km s$^{-1}$ and we estimate (supplementary material) that the average energy flux is ~$10^3$ W m$^{-2}$ in the spicule and ~300 W m$^{-2}$ in the lower corona, which is comparable to observations*(8,9)*. Dissipation of such waves, e.g., through resonant absorption, is not properly treated by our simulation because it does not include the small spatial scales on which such mode-coupling and subsequent Kelvin Helmholtz instability vortices occur, but dissipation can also lead to substantial heating of the outer atmosphere(11,27-29).

This work is supported by NASA, the Research Council of Norway and the European Research Council. IRIS is a NASA small explorer mission developed and operated by LMSAL with mission operations executed at NASA Ames Research center and major contributions to downlink communications funded by ESA and the Norwegian Space Centre. The Swedish 1-m Solar Telescope is operated on the island of La Palma by the Institute for Solar Physics of Stockholm University in the Spanish Observatorio del Roque de los Muchachos of the Instituto de Astrofísica de Canarias. We gratefully acknowledge support by NASA grants NNX11AN98G, NNM12AB40P, NNH15ZDA001N-HSR, NNX16AG90G, and NASA contracts NNM07AA01C (Hinode), and NNG09FA40C (IRIS). This research was supported by the Research Council of Norway and by the European Research Council under the European Union's Seventh Framework Programme (FP7/2007-2013) / ERC Grant agreement nr.

291058. The simulations have been run on clusters from the Notur project, and the Pleiades cluster through computing project s1061 from NASA's High End Computing Capability (HEC). We thankfully acknowledge the computer and supercomputer resources of the Research Council of Norway through grant 170935/V30 and through grants of computing time from the Programme for Supercomputing. The IRIS observations are online and freely accessible (http://iris.lmsal.com/data.html). Level 3 data, which includes both SST and IRIS observations used for Figure 4, are accessible at the IRIS webpage (http://bit.ly/2mnGVYg). The RH code used to calculate the synthetic observables is freely accessible at http://github.com/ITA-Solar/rh and documented (ITN 35, 36 and 37 documents in http://iris.lmsal.com/modeling.html, *(46-47)*). Likewise, the code that calculates the optically thin radiation is available in the SolarSoftWare (SSW) package, under the IRIS library (SSW/IRIS). This software is documented (ITN34, http://iris.lmsal.com/modeling.html).  The initial conditions of the radiative MHD simulation are available upon request from the authors.

Supplementary Materials
www.sciencemag.org
Materials and Methods
Figs. S1, S2, S3, S4, S5 and S6
Table S1
References (30-69)
Caption for Movies S1-S5.

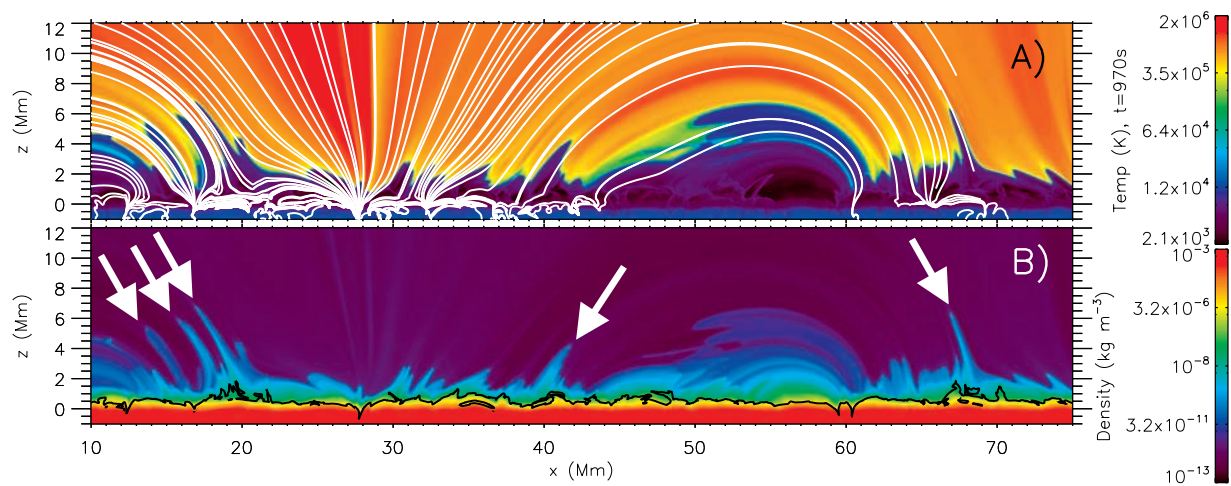

**Figure. 1**. **Spicule-like features** (indicated by arrows) **occur in our radiative MHD simulations**. They appear as dense, cool intrusions in the hot corona, originating from the boundaries of the strong magnetic regions. Temperature (A) and density maps (B) are shown on a logarithmic color map with white magnetic field lines (A) and black (B) contours where plasma β=1.

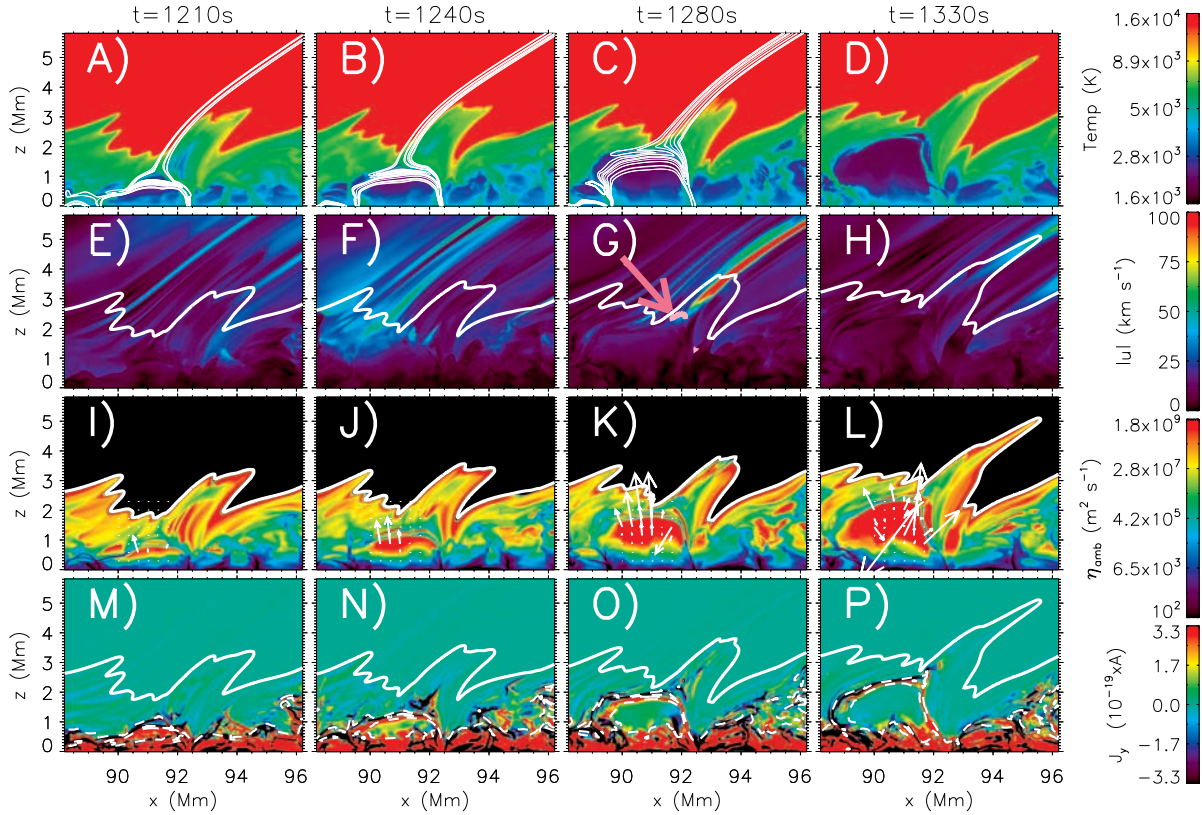

**Figure. 2. Spicules form when strong magnetic tension is diffused into the upper chromosphere where its release drives strong flows, heating, and Alfvén waves.** A time series is shown of temperature, absolute velocity, ambipolar diffusion and current perpendicular to the plane. The thick white contour is at $10^5$ K (E-P). The region of strong tension is illustrated in panels A-C with white magnetic field lines. White arrows in panels I-L show ambipolar velocities leading to upwards diffusion of the high-tension region. Strong ambipolar diffusion in the expanding emerging flux bubble at $t$=1280 s concentrates the perpendicular current at the edges of the bubble, amplifying the tension. Pink contours (G) show locations of strong plasma pressure gradient (driving upwards spicular flows) resulting from the release of magnetic tension (pink arrow).

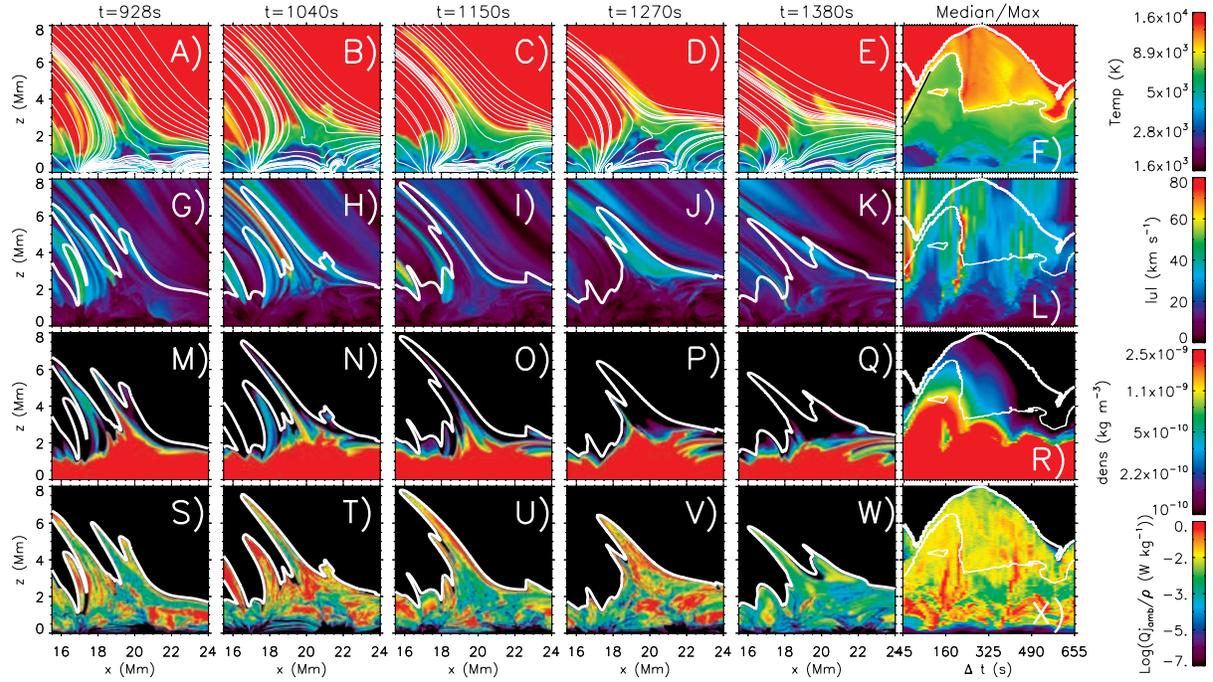

**Figure. 3. Heating of spicules from dissipation of electrical currents through ambipolar diffusion.** Time series of maps of temperature, absolute velocity, density and Joule heating per particle from ambipolar diffusion are shown from top to bottom respectively. The median temperature (F), maximum velocity (L), median density (R) and median Joule heating per particle from ambipolar diffusion (X) within the spicule ($<10^5$ K) are shown as a function of time and height. Magnetic field lines are in white (A-E), and the thick white contour is at $10^5$ K in the three bottom rows. The rightmost column includes a thin contour at $8\times10^3$ K.

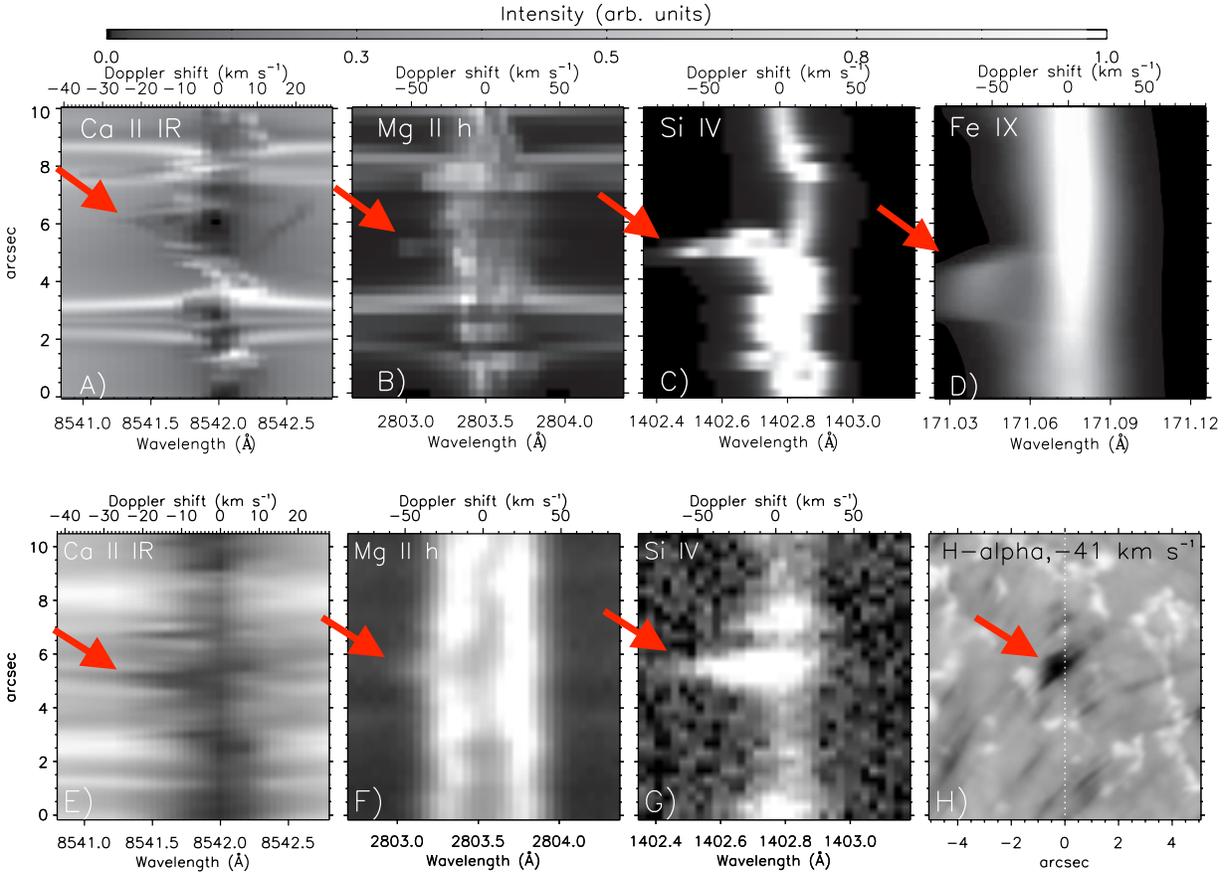

**Figure 4: Synthetic observations of a simulated spicule compared with observations** from the Swedish 1-m Solar Telescope (SST) and the Interface Region Imaging Spectrograph (IRIS). Top row shows wavelength-space plots of synthetic observations of Ca II 8542 Å (middle chromosphere), Mg II h 2803 Å (upper chromosphere), Si IV 1403 Å (TR) and Fe IX 171 Å (corona). The spicular signal appears as a blueward excursion around x=3-6" (red arrows). The signals are offset spatially because the spicule is inclined from the vertical. Bottom row shows a disk observation of a spicule (or RBE) in Ca II 8542 Å (SST), Mg II h 2803 Å (IRIS) and Si IV 1403 Å (IRIS), and a map of the blue wing (-41 km s$^{-1}$) of Hα 6563 Å (SST). The white dotted line in (H) indicates the location of the IRIS slit. The spatial range (A-D) corresponds to the region shown in Figure 3 at t=929s.

# 1. Material and Methods

## 1A. Ion-neutral effects

Partial ionization effects play a major role in many astrophysical features such as the interaction between the heliosphere and local interstellar medium *(31)*, protoplanetary disks *(32),* the ionosphere, the magnetic cusp and planetary magnetopauses and magnetospheres *(33-34),* accretion disks *(35),* and also in solar prominences *(36)* among others. We investigate the ion-neutral interaction effects in spicule formation by modeling the solar atmosphere using the Bifrost code *(37)*. This software solves the full MHD equations with non-grey *(38)*, non-Local Thermodynamic Equilibrium (non-LTE) radiative transfer, including coherent scattering and thermal conduction along the magnetic field. The numerical methods implemented in the code are described and tested in great detail by *(37, 39, 40)*.

In order to implement ion-neutral interaction effects we take into account that the code explicitly solves the MHD equations on a Cartesian staggered mesh. We implement partial ionization effects in the Bifrost code by adding two new terms to the induction equation: a Hall term (the separation of electrons and ions and the collision between neutrals and ions lead to a generation of electric current perpendicular to the magnetic field) and a term describing ambipolar diffusion *(41-43)*. Inserting the generalized Ohm's law into the induction equation we find:

$$\partial \boldsymbol{B}/\partial t = \nabla \times [\boldsymbol{u} \times \boldsymbol{B} - \eta \boldsymbol{J} - \frac{\eta_{\text{Hall}}}{|\boldsymbol{B}|} \boldsymbol{J} \times \boldsymbol{B} + \frac{\eta_{\text{Amb}}}{|B^2|}(\boldsymbol{J} \times \boldsymbol{B}) \times \boldsymbol{B}] \quad \text{(S1)}$$

where $\boldsymbol{B}$, $\boldsymbol{J}$, $\boldsymbol{u}$, and $\eta$, $\eta_{\text{Hall}}$, $\eta_{\text{Amb}}$ represent the magnetic field, the current density, the velocity field, ohmic diffusion, the Hall term, and the ambipolar diffusion, respectively. The generalized Ohm's law can be reformulated as follows:

$$\partial \boldsymbol{B}/\partial t = \nabla \times [\boldsymbol{u} \times \boldsymbol{B} - \eta \boldsymbol{J} + \boldsymbol{u}_{\text{Hall}} \times \boldsymbol{B} + \boldsymbol{u}_{\text{Amb}} \times \boldsymbol{B}] \quad \text{(S2)}$$

where the Hall *velocity* is $\boldsymbol{u}_{\text{Hall}} = -\eta_{\text{Hall}} \boldsymbol{J}/|\boldsymbol{B}|$ and the ambipolar velocity is $\boldsymbol{u}_{\text{Amb}} = \eta_{\text{Amb}} \boldsymbol{J} \times \boldsymbol{B}/|B^2|$ *(44)*. In order to suppress numerical noise, high-order artificial diffusion is added both in the forms of viscosity and magnetic diffusivity *(37)*. The implementation of these two terms in the induction equations is extensively described and tested in *(43)*. The ion-neutral cross sections are listed in *(45)*.

The simulated atmosphere spans from the upper layers of the convection zone (2.5 Mm below the photosphere) to the corona (40 Mm above the photosphere). The horizontal domain spans 96 Mm. Convective motions perform work on the magnetic field and introduce magnetic field stresses in the corona. This energy is dissipated and creates the corona self-consistently as the energy deposited by Joule heating is spread through thermal conduction *(46)* whereby the temperature reaches up to two million Kelvin (Fig. 1A). The spatial resolution is uniform along the horizontal axis (14 km) and non-uniform in the vertical direction, allowing a smaller grid size in locations where it is needed, such as in the photosphere and in the transition region (~12 km). Experiments with our numerical simulations show that a spatial resolution of 40 km or better is required to properly resolve the small-scale structuring in the magnetic field, flow-field and current. These play a critical role in the spicule formation and ambipolar diffusion. With the resolution used in our simulations, i.e., ~12 km, the ambipolar diffusion is larger than the artificial diffusion (by 3 to 5 orders of magnitude) in extended regions in the chromosphere as shown in Figure S1, allowing us to properly capture these effects.

The initial magnetic field contains two plage-like regions of high magnetic field strength and opposite polarity. These are connected by magnetic loops that are up to ~50 Mm long (Fig. 1). The mean absolute value of the magnetic field in the photosphere is ~190 G. After this initial setup we allow the model to relax for 40 min of simulated time to ensure that transients have propagated out of the numerical domain. No magnetic flux is injected through the bottom boundary.

## 1B. Synthetic profiles

In order to compare our results with observations, we have synthetized spectral profiles of the chromospheric Ca II 8542 Å and Mg II h 2803 Å lines, and EUV lines that form at transition region and coronal temperatures: Si IV 1402 Å (formed roughly at $8 \times 10^4$ K) and Fe IX 171 Å (formed at ~$9 \times 10^5$ K).

The Ca II 8542 Å and Mg II h 2803 Å diagnostics have been calculated using the RH code *(47, 48)*, which solves the non-LTE radiative transfer problem in 1D on a column-by-column basis (1.5D approximation), for each snapshot of the simulation (as viewed from above). This 1D treatment is a good approximation for radiative transfer in the Ca II lines, except at the very line core *(49)*. Mg II h and k could suffer from 3D effects in the peak intensity, but not the Mg II h and k wings *(50)*. For Ca II, a 5-level plus continuum calcium model atom was employed, while for Mg II we used a 10-level plus continuum model atom *(51)*. The radiative transfer calculations were carried out in non-LTE, allowing for partial redistribution (PRD) in the Mg II h & k lines (using the hybrid angle-dependent PRD recipe of *(52)*), and complete redistribution for all other transitions.

The emission for transition region and coronal EUV lines is calculated following the methods described in *(26)*, using the optically thin approximation under ionization equilibrium conditions. To synthesize the plasma emission we use CHIANTI v.7.0 *(53, 54)* with the ionization balance chianti.ioneq, available in the CHIANTI distribution. We synthesized observations for so called 'photospheric' abundances *(55)*.

For comparison with IRIS and SST observations, the synthetic profiles have been convolved spatially to 0.166" for Ca II and 0.33" for Mg II and Si IV. All three profiles are resampled to 0.33" per pixel. The profiles were spectrally convolved to 0.05054 Å,

0.02585 Å and 3 km s$^{-1}$ FWHM, and resampled to 0.05092 Å, 0.02585 Å, and 2.5 km s$^{-1}$ for Mg II, Si IV, and Ca II, respectively.

## 1C. Observations

The observations shown in Figure 4 were taken at the SST*(20)* and IRIS*(21)* on 26-June-2015 from 07:09 UTC to 09:40 UTC. The field-of-view was centered on solar (*x,y*) coordinates of (-423", -259") which was in the middle of a strong network area.

The Ca II 8542 Å (panel F in Fig. 4) and Hα 6563Å (panel G in Fig. 4) spectral scans were obtained with the CRisp Imaging SpectroPolarimeter (CRISP) Fabry-Perot Interferometer *(56)* at the SST. CRISP scans had a temporal cadence of 16 seconds, while the native spatial sampling of the CRISP data is 0.057" per pixel and the diffraction limit is 0.18". In figure 4 the SST spectral resolution has been interpolated to IRIS spectral resolution. The CRISP data were processed following the CRISPRED data reduction pipeline *(57),* which includes Multi-Object Multi-Frame Blind Deconvolution image restoration *(19,58).* The alignment with the IRIS data was done through cross-correlation of the Ca II 8542 Å wing and the IRIS SJI 2796 channel.

The Mg II h 2803 Å (Fig. 4F) and Si IV 1402 Å (Fig. 4G) spectra were obtained with IRIS running an observing program labeled OBS-ID 3630105426. This program comprises an 8-step raster with continuous 0.33" steps of the 60" spectrograph slit covering a spatial area of 2.3" x 60" at a temporal cadence of 25 seconds. To increase signal-to-noise the IRIS data was summed on board (2x2 pixel binning) so that the original IRIS resolution of 0.166" spatial sampling was reduced to 0.33" spatial sampling, while the spectral resolution was reduced from 3 km s$^{-1}$ spectral sampling to 6 km s$^{-1}$ sampling.

# 2 Supplementary Text

## 2A. On the 2.5D limitation of the simulations

The simulations we present are by necessity in 2.5D. Currently available computational facilities are insufficient to perform 3D simulations with the required size of the numerical domain, and the necessary physical complexity and resolution required for properly treating the effects of ion-neutral collisions through the Generalized Ohm's law.

Three of our simulations address the limitations inherent to 2.5D simulations and show that the mechanism that is at the core of this paper is not an artifact of the 2.5D simulations. Table S1 illustrates the properties of these simulations:

**Model A** was used by *(12,15)*. This model covers a rather small 3D domain with radiation, thermal conduction, and strong flux emergence introduced at the bottom boundary. This model does not include ambipolar diffusion.

**Model B** is the main focus of this paper, i.e., large scale high resolution 2.5D radiative MHD including ambipolar diffusion. This model does not include flux emergence introduced at the bottom boundary. The interaction between strong and weak fields that plays a key role in the spicule formation mechanism is facilitated by ambipolar diffusion and occurs self-consistently from the interaction between the magnetic field and convective motions.

**Model C** is identical to Model B, but without ambipolar diffusion. Model C is the only model that does not produce fast type II spicules.

Figure S2 shows that ambipolar diffusion is a crucial ingredient in driving fast spicules. The figure shows the temperature and velocity in models B and C. Model C, which lacks ambipolar diffusion, shows slow (<40 km s$^{-1}$) spicules that reach heights of less than 5 Mm and that are not heated to transition region temperatures, while Model B shows fast (>40 km s$^{-1}$) spicules that reach heights up to 10 Mm. This confirms that the ambipolar diffusion is crucial to generate type II spicules. In other words, the mechanism for spicule formation that we describe in the current paper is not an artifact of the 2.5D assumption.

Model B, the main focus of this paper, is able to produce a large number of fast spicules. It also produces chromospheric and transition region observables that match with observations from SST and IRIS (e.g., RBEs), and that reproduce the observed thermal evolution.

Model B differs from previous results (e.g., Model A by *(12, 15) and (59-60) among others)*:

- Model A produced at most two spicule-like features.
- The synthetic observables and thermal evolution of the one or two spicule-like features in model A are not compatible with observations *(13)*.
- Model A is a single-fluid MHD simulation, i.e., lacks ambipolar diffusion. Because ambipolar diffusion facilitates expansion of magnetic flux into the atmosphere, model A includes less frequent expansion of flux into the atmosphere, and thus only rarely gives rise to the conditions necessary for spicule formation (as described in the current paper).
- Model A only produced spicules when the strong flux emergence that was introduced at the bottom boundary occasionally mimics the conditions that self-consistently (i.e., without introduction of flux emergence at the bottom boundary) occur in model B.
- Spicules occur much more frequently in model B, and similar to observations, occur preferentially in the vicinity of strong magnetic flux in network and plage regions.

The results of model B thus differ from those of model A, suggesting that ambipolar diffusion is the missing ingredient required for spicule formation.

Both models A and B show that the mechanism we describe is not limited to 2.5D. Model A shows that the release of magnetic tension from this type of magnetic field configuration leads to a collimated jet, also in 3D. This is confirmed by model B, which shows that the strong flows along the magnetic field are highly collimated jets (figure S6 and movie S5): the jets do not suffer any expansion along the x axis. In both models A and B, the release of the magnetic tension squeezes the plasma in a very localized region (e.g., see the pressure increase in Figure 2). The ambipolar diffusion in model B further concentrates the driving forces into a smaller region.

In strong field regions such as plage and network the expansion of the field with height is reduced, thus presenting a more 2.5D type environment, and allowing the forces to remain more concentrated than expected from expansion to 3D.

## 2B. Ambipolar flows

One way to visualize the effects of ambipolar diffusion is by using expression (S2). Ambipolar diffusion moves the magnetic field lines in a direction (and amplitude) dictated by the ambipolar velocity, thereby relaxing the magnetic tension of the field lines. This ambipolar velocity can be interpreted as the velocity drift between ions and neutrals. Figure S3B shows the ambipolar velocity in the cold pockets. Without ambipolar diffusion, horizontal magnetic fields require high field strengths to become buoyant in the sub-adiabatic photosphere *(17,61)*. Under such conditions, weak field regions have difficulty expanding into the atmosphere.

When ambipolar diffusion is large enough, e.g., in very cold photospheric pockets, it allows the magnetic field to propagate through the photosphere into the upper atmosphere (Figure S3B). In addition, the current (Fig. S3D) is advected to the exterior of the cold pocket due to the ambipolar velocity field. Despite the fact that the magnetic field expansion tends to push mass into the upper chromosphere as the field expands, ambipolar diffusion reduces this effect since the magnetic field is not fully frozen in with the plasma *(62-65)*. This allows the horizontal magnetic field to rise to greater heights. Eventually the tension of the highly curved field lines on the right hand side of the spicule in Fig. S3 squeezes the plasma against the pre-existing ambient field above and thereby produces fast flows (up to 100 km s$^{-1}$).

## 2C. Other drivers

In the main text we described one mechanism that can lift the highly bent magnetic field lines from the photosphere (high plasma β) to the chromosphere (low plasma β). In addition to ambipolar diffusion, there are other processes that can lift bent magnetic field into the upper layers of the solar atmosphere.

Previous work has showed that small-scale flux emergence sometimes has enough buoyancy to go through the sub-adiabatic photosphere, which in numerical models has been found to occasionally drive a spicule *(15)*. However, without ambipolar diffusion, this process does not appear to produce spicules in a ubiquitous manner. In addition, the few spicules that are produced do not match all of the observed properties.

In our current model, we have also found another mechanism that can lead to more ubiquitous formation of spicules that do match the thermal evolution seen in observations. Convective motion at the boundaries of granules causes buffeting of the horizontal overlying field lines producing magneto-acoustic waves propagating almost parallel to the photosphere. Figure S4 shows an example of such an event. The velocity undergoes a jump, transverse to the magnetic field lines that propagates almost parallel to the photosphere. Current perpendicular to the magnetic field lines is associated with this wave, i.e. it is a magneto-acoustic shock (Fig. S4P-R). This shock leaves behind a small cold pocket where the ambipolar diffusion becomes important (Fig. S4M-N). The combination of the shock passing through the highly inclined field lines and the ambipolar diffusion in the cold pocket behind the shock amplifies and ultimately releases the tension at a location just below the future spicule (Fig. S4R-S). The magnetic field reconnects somewhere in the photosphere on the left hand side of the spicule in Fig. S4 (around x=18 Mm). However, the driver of the spicule comes from the opposite side (the shock is coming from x=23 Mm). The cold pocket in this case is a consequence of the shock traveling almost horizontally, whereas in the example described in the main text, the pocket is due to the expansion of the magnetic field. The magnetic field lines that drive the acceleration of the spicular plasma connect the corona to field lines that are nearly horizontal just above the photosphere. For the case shown in figure S4, the highly bent magnetic field suffers an increase of magnetic tension due to the combination of the shock and the ambipolar diffusion behind the shock which tends to concentrate tension in an even thinner layer. Finally, the release of magnetic tension increases the plasma velocity to Alfvénic speeds.

In both of these cases the ambipolar diffusion is large enough in the lower chromosphere and/or upper photosphere to play a crucial role in driving the spicule. For the driver described in the main text, once the magnetic field goes through the photosphere, the magnetic field expands drastically, producing a cold pocket characterized by low temperatures in the lower chromosphere. The expanding magnetic field carries a region of strong tension on the right hand side of the cold pocket due to an earlier reconnection in the photosphere. In contrast, in the example described here, convective motion perturbs the overlying horizontal magnetic field producing waves traveling along the horizontal magnetic field (at around x = 24 Mm in Fig. S4F-I).

## 2D. Waves driven by the spicules

The simulated spicules drive both transverse and acoustic shocks/waves. The whiplash effect or release of magnetic tension drives both the strong flows along the magnetic field lines and transverse waves. The latter can be seen in maps of the transverse velocity (uy), perpendicular to the plane of this 2.5D simulation (Figure S5). This figure shows that the spicule can drive transverse waves with relatively short periods (~30-150s).

Figure S6 (and Movie S5) shows that the transverse waves described above, which drive the spicule, are converted to both acoustic shock waves (Fig. S6I-J) and transverse waves (Fig. S6N). The enhancement of the magnetic tension due to the horizontal traveling wave along the bent magnetic field decreases the value of the plasma β. This can be seen by comparing the β=1 contour near the shock in Fig. S6L-M. this results in release of the magnetic tension and leads to longitudinal and transverse waves.

The transverse wave propagates for tens of seconds into the corona at Alfvénic speeds. Its peak transverse velocity amplitude is of order 50 km s$^{-1}$. Taking into account the local Alfvén speed (~300 km s$^{-1}$ at z=4 Mm inside the spicule and ~2000 km s$^{-1}$ at z~10 Mm in the corona), densities (~1.3 10$^{-9}$ kg m$^{-3}$ at z=4 Mm and ~10$^{-12}$ kg m$^{-3}$ at z=10 Mm), and transverse speeds (~50 km s$^{-1}$) we estimate that the energy flux of transverse waves, i.e., the kinetic energy of the transverse velocity multiplied by the Alfvén speed, reaches peak values of ~25 kW m$^{-2}$ in the spicule and ~3 kW m$^{-2}$ in the lower corona (Fig. S6C-E).

The whiplash effect leads to transverse waves in 2.5D simulations like ours, but would lead to both torsional and kink waves on the Sun *(11)*. The values given above represent the largest found during the formation of these waves. Averaged over the whole spicule lifetime, i.e. the time between the initial upflows and the return of the plasma to its initial chromospheric heights (several minutes), we find that the mean energy flux is ~1 kW m$^{-2}$ in the spicule, ~0.3 kW m$^{-2}$ in the corona, with transverse amplitudes of ~20 km s$^{-1}$. These values are compatible with observational estimates *(8, 9)*. The dissipation of such waves, e.g., through resonant absorption, can also lead to significant heating of the outer atmosphere *(8,11,27-29)*.

The acoustic wave is supersonic, because the gradient of pressure that drives the waves is triggered by the release of magnetic tension and thus can be very large in a low plasma beta environment like the upper chromosphere.

## 2E. Thermal and magnetic evolution of the spicule

Despite the large expansion of the plasma as the spicule forms, the temperature of the spicule increases, i.e., there is significant heating that compensates for the cooling due to adiabatic expansion. This is related to the fact that these spicules carry current parallel to their axis. Inside the spicule, this current is dissipated through ambipolar diffusion that heats the spicular plasma to transition region temperatures, despite the strong expansion (Fig. 3S-X and movie S3). In the vicinity (corona) of the simulated spicules the dissipation is due ohmic diffusion. Ambipolar diffusion is thus larger at the spicule tops since the ion-neutral collision frequency is smaller there. This occurs because of the rapid expansion of the spicule and the subsequent decrease of density. Because the density is lower here than at deeper regions in the chromosphere, ambipolar heating is more effective (Fig. 3M-R). These spicules carry current and waves transverse to the magnetic field, which are partially dissipated due to the ambipolar diffusion. This temperature evolution is in accordance with the limb observations of type II spicules that disappear in Ca II (within a couple of minutes) but remain visible in Mg II or transition regions lines *(5,66)* (for almost 10 minutes). The simulated spicule density decreases substantially during their evolution (Fig. 3N-O), which can also help explain the reduced visibility in Ca II H after the initial acceleration phase.

Despite the significant heating in these spicules, with a fraction of the chromospheric plasma heated to transition region temperatures (via ambipolar diffusion) and some even to coronal temperatures (via ohmic diffusion), much of the chromospheric plasma falls back to the chromosphere at chromospheric temperatures. There is strong heating also in the vicinity of the spicules, both above and in the immediate surroundings. For example, the loops associated with the spicule are also strongly heated by Joule dissipation of currents. In the simulation, the magnetic dissipation in the corona is via artificial diffusion, whereas in the Sun the current could be dissipated by ohmic diffusion. The deposited energy propagates along the magnetic field lines through thermal conduction. Movie S4 shows that the loops connected to the spicules are heated to temperatures of up to 2 MK. The currents associated with the spicule are not only located at the footpoints, but also throughout the spicule body and its immediate surroundings (Fig. S3). This means that heating occurs not only in plasma that was originally at chromospheric temperatures (spicule plasma), but also in plasma that was originally at transition region temperatures, or even at coronal temperatures. The heating of the plasma to coronal temperatures in association with spicule acceleration is thus not only caused by spicule plasma being heated to coronal temperatures. The heating associated with spicular acceleration varies from event to event.

The last instances of the spicule in figure 3 and movie S3 reveal a misalignment of the magnetic field lines with the thermal structure of the spicule. The misalignment is due to the ambipolar velocity, which moves the magnetic field in a different direction than the plasma flow (see eq. S2). Consequently, during the falling phase of the spicule it also drifts sideways. This is further investigated in *(65)*.

## 2F. Comparison to observations

Simulated spicules occur frequently in the vicinity of enhanced network and plage region in accordance with observations (Fig. 1).

For limb measurements the motions of the simulated spicule are very similar to those in observations (Figure 3F):

- Both observations and simulations show parabolic motions in height vs. time plots with similar lifetimes of 5-10 minutes for the full evolution (i.e., from the initial strong upflows until the return to chromospheric heights at the end of the lifetime).
- Observers typically report the maximum velocity along the spicules, which are usually observed during their initial stage *(3, 4)*. To compare observed velocities with our simulations, we focus on the initial rapid evolution of the height of the transition region (i.e., the top of the chromospheric spicule), which in this example moves up by about 2000 km in height during 60 s, i.e., about 33 km s$^{-1}$. However, this is the vertical motion of the spicule, not the motion along the spicule (the latter is usually reported for observations). Since this example spicule has an inclination of 40 degrees from the vertical, the velocities along the spicule, during its violent initial phase, are of order 40-45 km s$^{-1}$. This agrees well with the observed range of maximum velocities of type II spicules *(4)*.

For on-disk measurements the synthetic observables are in qualitative agreement with observations of RBEs, and the on-disk counterparts of type II spicules, with similar Doppler shifts and lifetimes. Doppler shifts from the top view are influenced by the upward velocity of spicules; the simulated RBE profiles have Doppler shifts that agree very well with the observations, both in Ca II 8542Å and Mg II k.

The simulated spicules thus have similar properties to observed type II spicules and their on-disk counterparts (RBEs). They are different from type I spicules (and their on-disk counterparts) in several ways. Type II spicules are faster, reach greater heights and include heating of plasma to TR temperatures (unlike type I spicules *(5,66)*).

The simulated spicules also have a coronal counterpart. The initial phase of the spicular signal in the Fe IX 171Å line, caused by the shock passing through, is rather faint (Figure 4). Taking into account the spatial, temporal and spectral resolution of current coronal spectrographs such as Extreme-ultraviolet Imaging Spectrometer *(67)* on board Hinode *(68),* it would be difficult to resolve this signal, but these events would cause a relatively faint blue-shifted secondary component at 1-20% of the total intensity (depending on the background intensity). This is in agreement with previous observations by *(69,24).* It provides support for the suggestion of a correlation between the spectral line asymmetries in coronal lines and chromospheric spicules *(24).*

-

# Figures

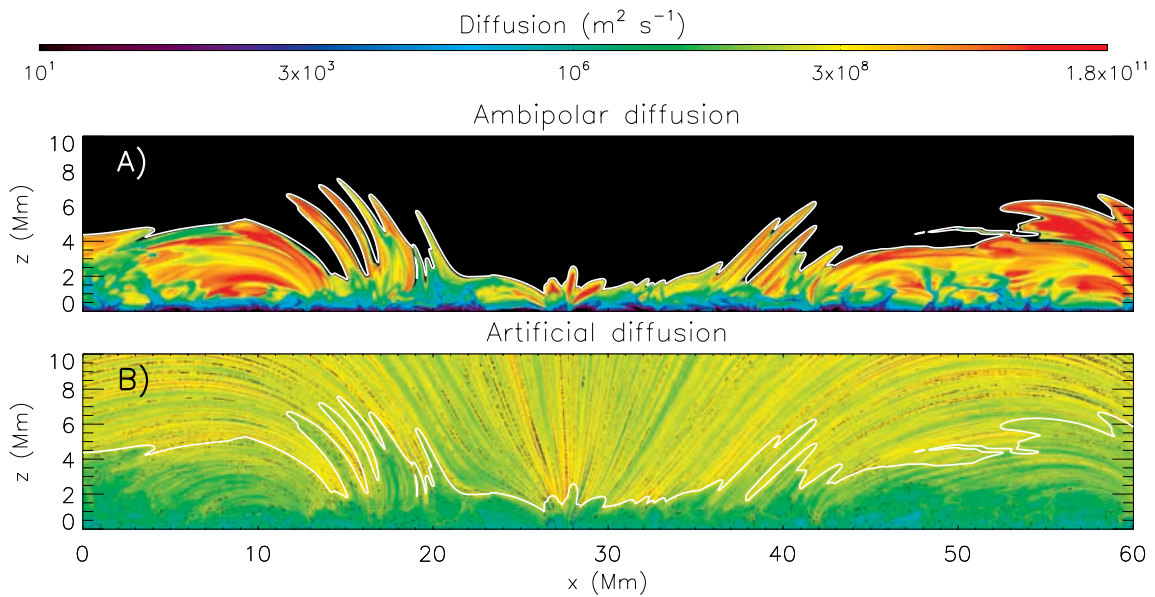

**Figure S1. The ambipolar diffusion** (A) **shows values that are 3 to 5 orders of magnitude larger than the artificial diffusion** (B) in extended regions of the chromosphere.

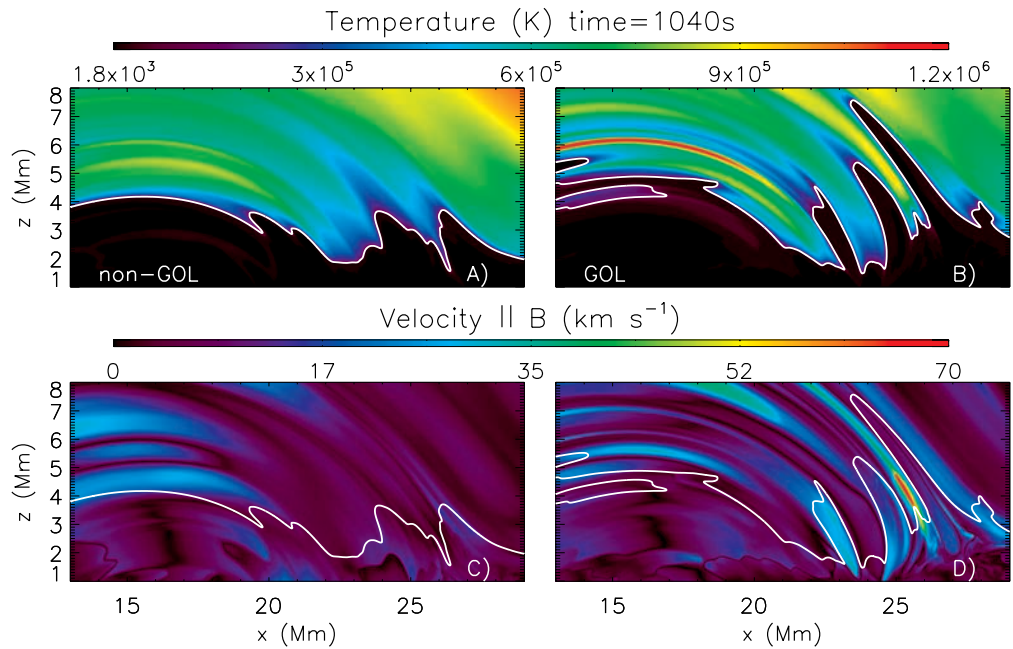

**Figure S2. Ambipolar diffusion is a crucial ingredient in driving fast and long spicules** in which plasma is heated beyond chromospheric temperatures. Temperature (A-B) and maps of the velocity parallel to the magnetic field (panels C-D) in models B (panels B,D), i.e., with ambipolar diffusion, and C, i.e., without ambipolar diffusion (panels A,C): panels A and C show slow (<40 km s$^{-1}$) spicules that reach heights of less than 5 Mm, while the panels B and D (including ambipolar diffusion) show fast (>40 km s$^{-1}$) spicules that reach heights up to 10 Mm.

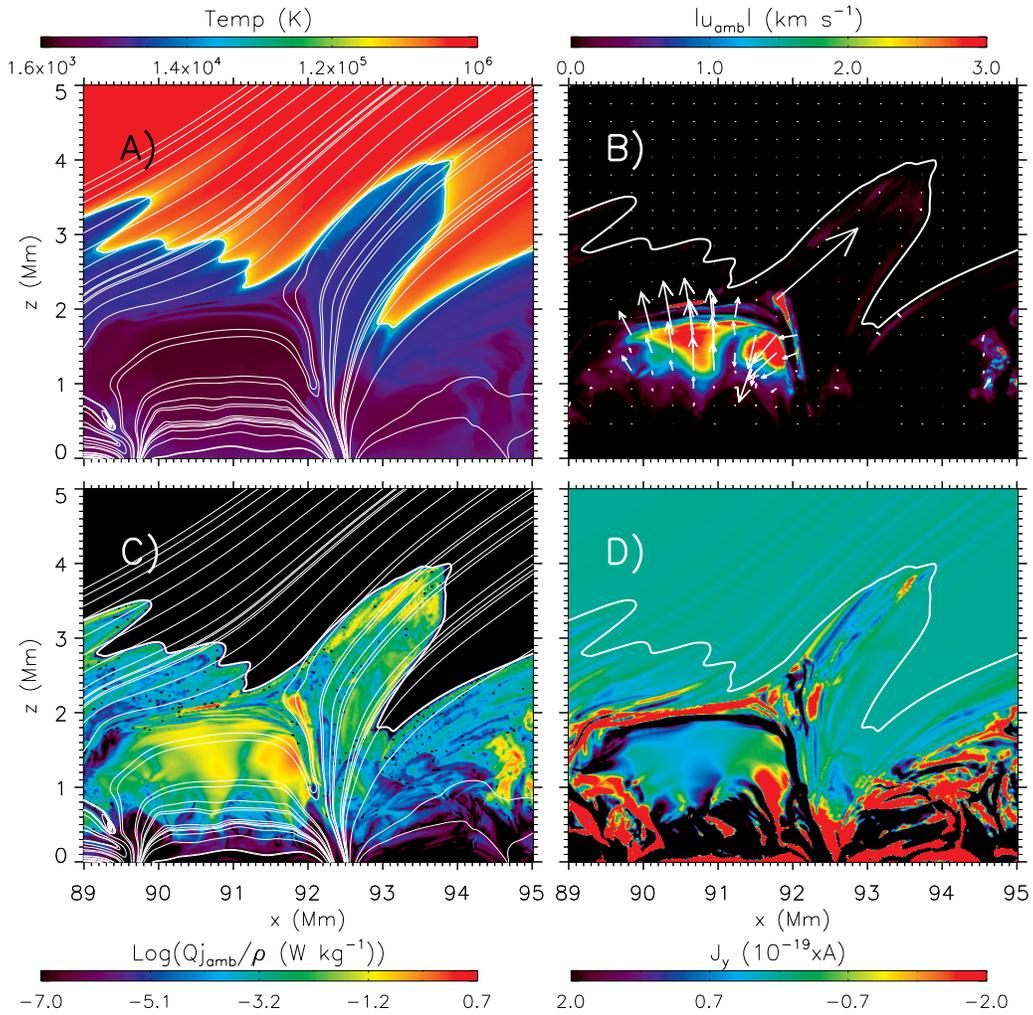

**Figure S3. Ambipolar velocity reveals that the horizontal field lines in the photosphere are diffused into the chromosphere and current is advected from the cold pocket to its boundaries, further concentrating the currents**. Maps of temperature, absolute ambipolar velocity, Joule heating from ambipolar diffusion, and the current perpendicular to the plane are shown in panels A-D respectively. Magnetic field lines are drawn in white on the panels A and C, and the thick white contour on panels B, C and D is at a temperature of $10^5$ K. The ambipolar velocity field is shown in panel B with white arrows.

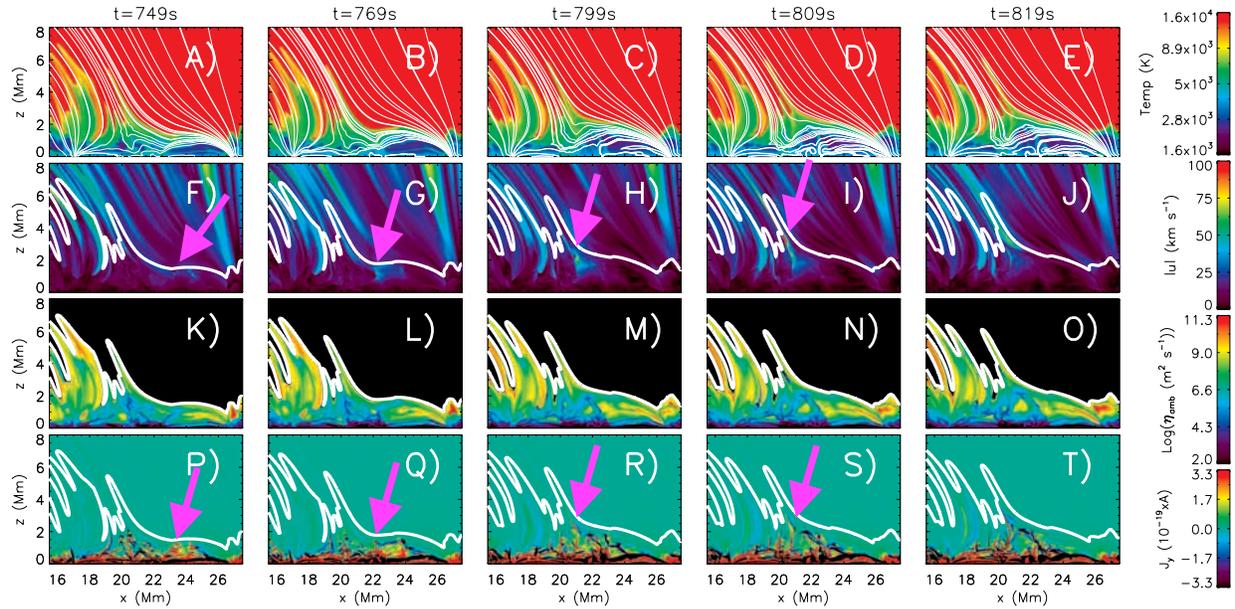

**Figure S4: The magnetic tension on the highly bent field lines can also be released by waves**. Time series of temperature maps with magnetic field lines in white are shown in panels A-E, of absolute velocity maps in panels F-J, of ambipolar diffusion maps in panels K-O, and of electric current density perpendicular to the plane in panels P-T. The white thick contours correspond to a $10^5$ K temperature. Arrows in panels F-I and P-S point to the wave, which allows the magnetic tension to be released to later drive the spicule.

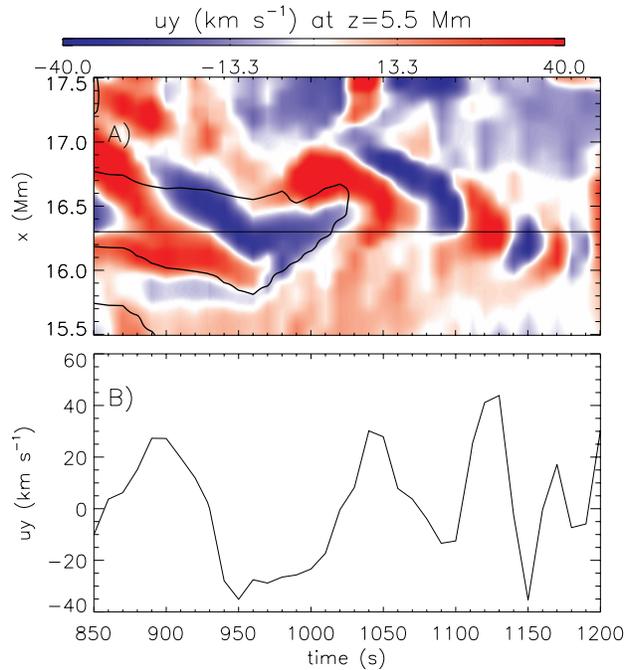

**Figure S5: The type II spicule driver generates transverse waves with relatively short periods (~30-150s).** The map (A) and the time plot (at x=16.3 Mm, B) of the velocity perpendicular to the simulated plane (uy) at a height z=5.5 Mm both show the oscillatory motion that occurs in association with the spicule that is formed at x~16.5 Mm. Note also that the oscillatory pattern appears to be displaced along the x-direction (with time) in an oscillatory fashion. The temperature at $10^5$ K is shown with black contours in panel A and indicates where the spicule is located. The thin horizontal line (A) is the location of the time-line plot (B).

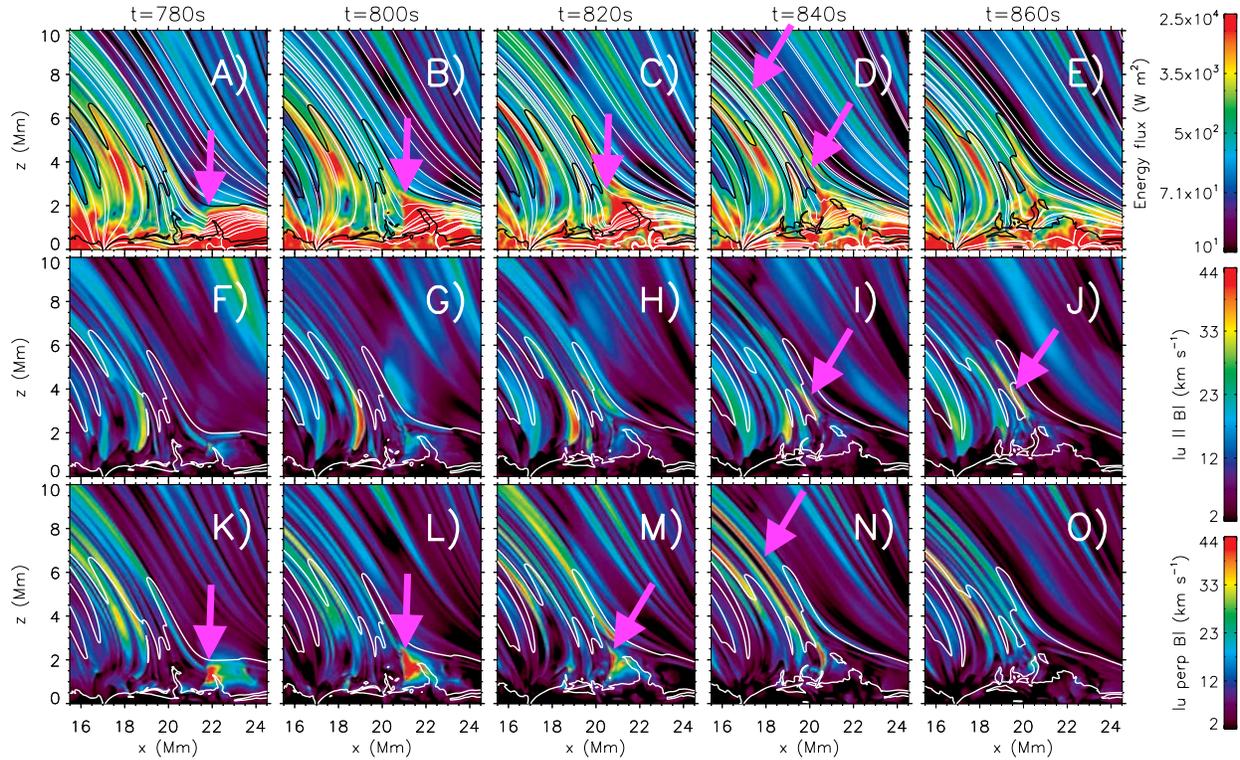

**Figure S6: Transverse waves propagating parallel to the photosphere drive supersonic flows along the bent field lines**. Time series of maps of the energy flux of transverse waves with magnetic field lines in white are shown in panels A-E, maps of absolute velocity parallel and perpendicular to the magnetic field in panels F-J and panels K-O respectively. The top contours correspond to a $10^5$ K temperature and the bottom contours correspond to plasma beta unity. Arrows in panels A-C and K-M point to the transverse wave, which allows the magnetic tension to be released, which later drives the spicule. This occurs through mode conversion of the transverse waves to acoustic waves that propagate parallel to the magnetic field (see arrows in panels I and J). In addition, transverse waves penetrate through the spicule into the corona (see arrows in panels D and N) (see also movie S5).

# Tables

| Model | Dimensions | Strong flux emergence | Generalized Ohm's law | Type II spicules | Thermal evolution | Reference |
|---|---|---|---|---|---|---|
| A | 3D | Yes | No | 1 or 2 | No | (12,15) |
| B | 2.5D | No | Yes | Many | Yes | This work |
| C | 2.5D | No | No | None | No | This work |

**Table S1**: Relevant properties of three different simulations, as follows: second column the number of dimensions of the models, third, fourth, and fifth columns list the presence of flux emergence, ambipolar diffusion, and type II spicules in the models respectively. The sixth column lists whether the models reproduce the observed thermal evolution of spicules. The last column lists the references for each simulation.

# Movies

**Movie S1**: The large-scale simulation reveals distinct regions where different processes occur. Spicule-like features naturally occur in these radiative MHD simulations. They appear as dense, relatively cool intrusions in the hot corona, originating from the boundaries of the strong magnetic field regions. The simulated spicules reach heights up to 10 Mm. Temperature (A) and density maps (B) are shown in a logarithmic scale. Magnetic field lines are drawn in white on the top panel and the height at which plasma $\beta=1$ is the thick red contour (A) and black contour (B).

**Movie S2**: Spicules form when regions of strong magnetic tension, generated at the solar surface from the interaction between vertical magnetic flux concentrations and horizontal fields, are diffused into the low plasma $\beta$ upper chromospheric region. This results in a violent release of the tension which drives strong flows and heating, and generates Alfvén waves. Time evolution of maps of temperature, ambipolar diffusion, absolute velocity and current perpendicular to the plane maps are shown in panels A-D. The temperature contour at $10^5$ K is the thick white contour. Magnetic field lines are drawn in white and they are advected taking into account the advection of the fluid and the ambipolar velocity. These lines are highlighting the region of strong magnetic tension. Strong ambipolar diffusion (t=1280s) in the expanding emerging flux bubble concentrates the perpendicular current at the edges of the bubble, amplifying the tension.

**Movie S3**: Heating of spicules from dissipation of electrical currents through ambipolar diffusion. The modeled spicule undergoes vigorous heating and reaches temperatures above $10^4$ K within two minutes. The total lifetime (as measured at higher temperatures)

is ~8 min, which is in agreement with observations. Time series of maps are shown for temperature (A), absolute velocity (C), density (E) and Joule heating per particle due to the ambipolar diffusion (G). The right column shows the median temperature (B), maximum velocity (D), median density (F) and median Joule heating per particle due to the ambipolar diffusion (H) within the spicule, i.e., over temperatures lower than $10^5$ K, as a function of height and time. Magnetic field lines are drawn in white (A). Temperature contours at $10^5$ K are shown in the three bottom rows. The right column also includes a temperature contour at $8 \times 10^3$ K. The red dashed lines in panel C shows the region where we calculated the medians or maximums for the right panels.

**Movie S4**: Hot loops (up to 2 MK) are formed in association with spicules. This can be seen with the temperature maps of the simulation with ambipolar diffusion and temperature contour at $10^5$ K shown in this movie.

**Movie S5**: Transverse waves propagating parallel to the photosphere drive supersonic flows along the bent field lines. Time series of maps of energy flux of transverse waves with magnetic field lines in white are shown in panel A, absolute velocity parallel and perpendicular to the magnetic field in panels B-C respectively. The top contours correspond to a $10^5$ K temperature and the bottom contours correspond to plasma beta unity. Magnetic field lines are drawn in white on panel A.